\documentclass[a4paper,12pt]{article}

\newcommand{\be}{\begin{equation}}
\newcommand{\ee}{\end{equation}}
\newcommand{\bea}{\begin{eqnarray}}
\newcommand{\nn}{\nonumber}
\newcommand{\eea}{\end{eqnarray}}

\begin{document}

\begin{titlepage}
\begin{flushright}
UB-ECM-PF-03/03
\end{flushright}
\begin{centering}
\vspace{.3in}
{\Large{\bf Casimir Effect, Achucarro-Ortiz Black Hole and the
Cosmological Constant}}
\\

\vspace{.5in} {\bf  Elias C.
Vagenas\footnote{evagenas@ecm.ub.es} }\\

\vspace{0.3in}

Departament d'Estructura i Constituents de la Mat\`{e}ria\\
and\\ CER for Astrophysics, Particle Physics and Cosmology\\
Universitat de Barcelona\\
Av. Diagonal, 647\\ E-08028 Barcelona\\
Spain\\
\end{centering}

\vspace{0.7in}
\begin{abstract}
 We treat the two-dimensional Achucarro-Ortiz black hole (also
 known as $(1+1)$ dilatonic black hole) as a Casimir-type system.
The stress tensor of a massless scalar field satisfying Dirichlet
boundary conditions on two one-dimensional ``walls'' (``Dirichlet
walls'') is explicitly calculated in three different vacua.
Without employing known regularization techniques, the expression
in each vacuum for the stress tensor is reached by using the
Wald's axioms. Finally, within this asymptotically non-flat
gravitational background, it is shown that the equilibrium of the
configurations, obtained by setting Casimir force to zero, is
controlled by the cosmological constant.
\end{abstract}

\end{titlepage}
\newpage

\baselineskip=18pt

\section{Introduction}
In the framework of quantum field theory in curved spacetime,
there is no natural definition of particles. Unfortunately, only
in exceptional cases the particle concept in curved spacetime
corresponds to the intuitive picture of subatomic physics
\cite{birrell}. Therefore, we are led to study  other observables
which are not globally defined, a thing that is obviously one part
of the problem with the particle definition. One of the most
interesting objects, if not the very first, is the stress (or
energy-momentum) tensor $T_{\mu\nu}(x)$. Furthermore, the interest
in explicitly calculating the stress tensor is augmented by the
presence of a gravitational background. The main reason is that
the role of the stress tensor is now twofold. It describes the
physical character of the quantum field at a spacetime point $x$
and it is also the source of gravity in this gravitational
background. There is a plethora of field theoretical procedures
\cite{birrell,fulling,rob,bryce,kummer}, known as regularization
techniques, for computing a finite and renormalized
$<T_{\mu\nu}>_{reg}$ such as the dimensional regularization
\cite{capper1,capper2,deser}, Green's function method
\cite{plunien,candelas}, heat kernel method \cite{greiner,gilkey},
zeta function regularization \cite{hawking1}, point-splitting
method \cite{davies1,davies2,christ}, Pauli-Villars regularization
\cite{vilenkin}. In this article, we are going to derive the exact
form of the stress tensor of a massless scalar field by
implementing some general properties of the renormalized stress
tensor known as Wald's axioms \cite{wald1,wald2} avoiding in this
way to employ any of the above-mentioned techniques.
\par
In 1948, H.B.G. Casimir \cite{casimir} was trying, in the
beginning, to calculate the Van der Waals force between two
polarized atoms. At the end, he was led to the problem of two
parallel conducting plates. He evaluated the attractive force
between the two plates and the electromagnetic energy which was
localized between the two conducting plates. The Casimir effect,
i.e. the disturbance to the electromagnetic vacuum induced by the
presence of two parallel conducting plates, is in contact with
laboratory physics \cite{bressi,bordag}. Nowadays, the
Casimir-type systems \cite{mostepanenko,bordag} are viewed as
tractable field theoretical models in which the general curved
spacetime formalism can be applied and sensible results can be
reached \cite{emilio1,emilio2,emilio3}.
\par
The scenario to be considered in our semiclassical analysis is as
follows. (a) The gravitational background will be the two
dimensional Achucarro-Ortiz black hole \cite{ana,martinez} which
is asymptotically an $AdS_{2}$ spacetime, (b) two one-dimensional
``walls'', separated by a distance L, are placed in the
aforementioned gravitational background and (c) the quantum field
whose stress tensor we are going to evaluate, will be a massless
scalar one satisfying Dirichlet boundary conditions on the
one-dimensional ``walls'' (``Dirichlet walls''). It is obvious
that the Achucarro-Ortiz black hole will be treated as a
Casimir-type system \cite{setare1,setare2,setare3,elias}.
\par
The paper is organized as follows. The next section is devoted to
 the presentation of Wald's axioms. In Section 3 and 4 we describe the
Achucarro-Ortiz and AdS$_{2}$ black hole geometries and we
calculate some of their geometrical quantities which would be
useful for the forthcoming analysis. In Section 5 the vacuum
expectation value of the stress tensor of the massless scalar
field in the Achucarro-Ortiz black hole geometry is explicitly
evaluated, respectively, in the Boulware vacuum (labeled by
$\eta$) \cite{boulware}, the Hartle-Hawking vacuum (labeled by
$\upsilon$) \cite{hartle,gibbons,israel}
 and the Unruh vacuum (labeled by $\xi$)
\cite{unruh}. The energy density, pressure, energy and the
corresponding force between the two ``Dirichlet walls'' are
specified. In Section 6, requiring the configurations to be in
equilibrium, the distance between the ``Dirichlet walls''  is
 seemed to be determined by the two-dimensional cosmological constant. Finally,
Section 7 closes with conclusions and prospects for future work.
\section{Wald's Axiomatic Analysis}
In mid-1970s there was a variety of techniques using complicated
mathematical devices for computing the stress tensors. There was
still the question of how to define a unique renormalized stress
tensor $<T_{\mu\nu}>$ purely by imposing physical requirements. R.
Wald proffered five ``axioms'' to be satisfied by the stress
tensors \cite{wald1,wald2}. The axioms, called from now on Wald's
axioms, are as follows
\begin{enumerate}
\item The expectation values of the energy-momentum tensor are covariantly conserved.
\item Causality holds.
\item In Minkowski spacetime, standard results should be obtained.
\item Standard results for the off-diagonal elements should also be obtained.
\item The energy-momentum tensor is a local functional of the metric, i.e. it depends only on
the metric and its derivatives which appear through the Riemann
curvature tensor.
\end{enumerate}
It should be remarked that recently there was a significant
generalization of the above-mentioned framework by R. Wald and S.
Hollands \cite{wald3}.\par\noindent Additionally, it must be noted
that in a clasical theory with a conformally invariant Lagrangian
the trace vanishes. However, in the corresponding quantized theory
the stress tensor may acquire a nonvanishing trace through
 renormalization (this is called conformal or trace anomaly)
\cite{capper1,capper2}. In two dimensions, the trace
$T^{\alpha}_{\alpha}$ can only be proportional to the Ricci scalar
$R$ of the theory \cite{deser,capri}. This is in agreement with
Wald's axioms.
\section{Achucarro-Ortiz Black Hole}
The black hole solutions of M. Ba$\tilde{n}$ados, C. Teitelboim
and J. Zanelli in ($2+1$) spacetime dimensions are derived from a
three dimensional theory of gravity \cite{jorge} \be S=\int dx^{3}
\sqrt{-g}\,({}^{{\small(3)}} R+2\Lambda)\label{btzaction} \ee with
a negative cosmological constant ($\Lambda=\frac{1}{l^2}>0$). The
corresponding line element is \be ds^2 =-\left(-M+\Lambda r^2
+\frac{J^2}{4 r^2} \right)dt^2 +\frac{dr^2}{\left(-M+\Lambda r^2
+\frac{J^2}{4 r^2} \right)}+r^2\left(d\theta
-\frac{J}{2r^2}dt\right)^2. \label{btzmetric}\ee There are many
ways to reduce the three dimensional BTZ black hole solutions to
the two dimensional charged and uncharged dilatonic black holes
\cite{ana}. The Kaluza-Klein reduction of the metric of the
($2+1$)-dimensional BTZ black hole (\ref{btzmetric}) yields a
two-dimensional line element
 \be ds^2 =-g(r)dt^2 +g(r)^{-1}dr^2
\label{metric1}\ee where \be g(r)=\left(-M+\Lambda r^2
+\frac{J^2}{4 r^2}\right)\label{metric2}
 \ee
with $M$ the ADM mass, $J$ the charge
 of the two-dimensional charged black hole, a U($1$) gauge field
\be A_{t}=-\frac{J}{2r^{2}} \ee and a dilaton field: \be \Phi= r.
\ee
\par \noindent
For the positive mass black hole spectrum with charge ($J\neq 0$),
the line element (\ref{metric1}) has two horizons \be
r^{2}_{\pm}=\frac{M\pm\sqrt{M^2 - \Lambda J^2} }{2\Lambda}
\label{horizon1} \ee with $r_{+}$, $r_{-}$ the outer and inner
horizon respectively. \par\noindent The Hawking temperature $T_H$
of the event (outer) horizon is \cite{kumar1} \bea T_H &=&
\frac{\sqrt{2\Lambda}}{2\pi}\frac{\sqrt{M^2-\Lambda J^2}}
{(M+\sqrt{M^2 -\Lambda J^2})^{1/2}}\nn\\
&=&\frac{\Lambda}{2\pi}\left(\frac{r_{+}^{2}-r_{-}^{2}}{r_{+}}\right)
\hspace{1ex}. \label{temp1}\eea \par\noindent The analytical
formulas for the nonvanishing  Christoffel symbols are \bea
\Gamma^{r}_{tt}&=&\frac{1}{2}\left(-M+\Lambda r^2 +\frac{J^2}{4
r^2}\right)\left(2\Lambda r-\frac{J^2}{2 r^3}\right)\label{aogamma1}\\
\Gamma^{r}_{rr}&=&-\frac{1}{2}\frac{\left(2\Lambda
r-\displaystyle{\frac{J^2}{2 r^3}}\right)}{\left(-M+\Lambda r^2
+\displaystyle{\frac{J^2}{4
r^2}}\right)}\label{aogamma2}\\
\Gamma^{t}_{rt}&=& \frac{1}{2}\frac{\left(2\Lambda
r-\displaystyle{\frac{J^2}{2 r^3}}\right)}{\left(-M+\Lambda r^2
+\displaystyle{\frac{J^2}{4
r^2}}\right)}\label{aogamma3}\hspace{1ex}.\eea The Ricci scalar is
given as \be
R(r)=-\left[2\Lambda+\frac{3J^{2}}{2r^{4}}\right]\label{aoricci}
\ee and therefore the nonzero trace o the stress tensor
corresponding to the Achucarro-Ortiz black hole takes the form \be
T^{\alpha}_{\alpha}(r)=-\left[\frac{\Lambda}{12\pi}+\frac{J^{2}}{16\pi
r^{4}}\right]\label{aotrace}\ee where we have used the expression
for the trace of a stress tensor in two dimensions
\cite{deser,capri} \be
T^{\alpha}_{\alpha}(r)=\frac{R(r)}{24\pi}\hspace{1ex}.\ee
\section{AdS$_{\bf{2}}$ Space}
The two-dimensional anti-de-Sitter geometry (AdS$_{2}$) can be
derived either by restricting the Achucarro-Ortiz black hole to
its spinless sector $J=0$ or by fixing the value of the dilaton
field which appears in the above-mentioned reduced theory
\cite{ana}. We adopt the first option and the resulting AdS$_{2}$
metric takes the form \be ds^2 =- g_{\mbox{\tiny{AdS}}}(r)dt^2
+g_{\mbox{\tiny{AdS}}}(r)^{-1}dr^2 \label{metric3}\ee where \be
g_{\mbox{\tiny{AdS}}}(r)=\left(-M+\Lambda r^2
\right)\label{metric4}\ee which has an horizon at \be
r_{H}=\sqrt{\frac{M}{\Lambda}}\hspace{1ex}. \label{horizon2} \ee
\par\noindent The temperature of the AdS$_{2}$ black hole is
\cite{kim}: \be T_H^{\mbox{\tiny{AdS}}} = \frac{\sqrt{\Lambda
M}}{2\pi}\; \hspace{1ex}\label{temp2} \ee
\par \noindent The analytical
formulas for the nonvanishing Christoffel symbols are \bea
\Gamma^{r}_{tt}&=&\Lambda r\left(-M+\Lambda r^2\right)\label{adsgamma1}\\
\Gamma^{r}_{rr}&=&-\frac{\Lambda r}{\left(-M+\Lambda r^2 \right)}\label{adsgamma2}\\
\Gamma^{t}_{rt}&=&\frac{\Lambda r}{\left(-M+\Lambda r^2
\right)}\label{adsgamma3}\hspace{1ex}.\eea The Ricci scalar is
given as \be R(r)=- 2\Lambda \label{adsricci}\ee and therefore the
nonzero trace of the Achucarro-Ortiz black hole takes the form \be
T^{\alpha}_{\alpha}(r)=-
\frac{\Lambda}{12\pi}\label{adstrace}\hspace{1ex}.\ee Using the
formula \be
T_{\mu\nu}^{\mbox{\tiny{AdS}}}=\frac{1}{\sqrt{-g}}\frac{\delta
\mathcal{L}_{grav} }{\delta
g^{\mu\nu}}\Bigg|_{g^{\mu\nu}=g^{\mu\nu}_{\mbox{\tiny{AdS}}}}\ee
the explicit expression for the stress tensor of  the
gravitational field of the AdS$_{2}$ space is easily calculated
\be T^{\mbox{\tiny{AdS}}}_{\mu\nu}= \left[\begin{array}{cc}
\displaystyle{\frac{r^2}{2}}&\displaystyle{0}
\\ \displaystyle{0}&\displaystyle{\frac{r^2}{2\left(-M+\Lambda r^2
\right)^2}}\end{array}\right]\label{adsstress}\hspace{1ex}.\ee
\section{Casimir Effect and Stress Tensor}
In this section a detailed expression for the renormalized stress
tensor of the massless scalar is obtained by enforcing the Wald's
axioms and using its trace. \par\noindent The starting point is
Wald's first axiom, i.e. that the conservation equation must be
fulfilled by the renormalized expectation value of the stress
tensor  $<T^{\mu}_{\quad \nu}>_{reg} \equiv T^{\mu}_{\quad \nu}$
\be T^{\mu} _{\quad \nu;\mu}=0 \label{conservation} \ee which
``splits'' in two equations  \bea
\frac{dT^{r}_{t}}{dr}+\Gamma^{r}_{rr}T^{r}_{t}
-\Gamma^{r}_{tt}T^{t}_{r}&=&0\\
\frac{dT^{r}_{r}}{dr}+\Gamma^{t}_{tr}T^{r}_{r}
-\Gamma^{t}_{rt}T^{t}_{t}&=&0 \eea and since
$T^{t}_{r}=-T^{r}_{t}$ and
$T^{t}_{t}=T^{\alpha}_{\alpha}-T^{r}_{r}$, we get  \bea
\frac{dT^{r}_{t}}{dr}+\Big[\Gamma^{r}_{rr}+\Gamma^{r}_{tt}\Big]T^{r}_{t}=0 \label{em1}\\
\frac{dT^{r}_{r}}{dr}+2\Gamma^{t}_{rt}T^{r}_{r}=
\Gamma^{t}_{rt}T^{\alpha}_{\alpha} . \label{em2} \eea Substituting
the Christoffel symbols (\ref{aogamma1} -\ref{aogamma3}) into
(\ref{em1}), (\ref{em2}) and solving them, we get respectively \be
T^{r}_{t}(r)=\frac{1}{g(r)} \delta \label{rttensor}\ee where\be
\delta=\alpha g^{\frac{3}{2}}(r) e^{-\frac{1}{4} g^{2}(r)}
\label{dconstant}\ee and \be
T^{r}_{r}(r)=\frac{1}{g(r)}\left[\beta + H_{2}(r)\right]
\label{rrtensor}\ee where \be
H_{2}(r)=\frac{1}{2}\int\limits^{r}_{r_{+}} \frac{d g(r')
}{dr'}T^{\alpha}_{\alpha}(r')dr' \label{rate} \ee and the
parameters $\alpha$, $\beta$ are constants of integration while
the point $r_{+}$ is where the outer horizon is placed. It can be
shown that $H_{2}(r)$ for the Achucarro-Ortiz black hole
background (\ref{metric1})-(\ref{metric2}) becomes  \be
H_{2}(r)=\frac{1}{96\pi}\left[2\Lambda r -\frac{J^{2}}{2
r^{3}}\right]^{2}-D \label{H21}\ee where D is constant \be
D=\frac{1}{96\pi}\left[2\Lambda r_{+} -\frac{J^{2}}{2
r_{+}^{3}}\right]^{2}\label{constant}\hspace{1ex}.\ee Now the
following limiting values of $H_{2}(r)$  are obtained from
(\ref{H21})
 \begin{eqnarray*}
\begin{array}{ll}
 if\hspace{0.5cm}
r \rightarrow r_{+} & then\hspace{0.5cm} H_{2}(r)= 0 \nn
\\ if\hspace{0.5cm} r \rightarrow +\infty  & then\hspace{0.5cm}
H_{2}(r)=\displaystyle{\frac{\Lambda^{2}}{24\pi}}r^{2} -
D\hspace{1ex}.\nn
\end{array}
\end{eqnarray*}
\noindent Therefore, using (\ref{rttensor}) and (\ref{rrtensor}),
we have the most general expression of the regularized stress
tensor in our gravitational background \be T^{\mu}_{\nu}=\left[
\begin{array}{cc}
 T^{\alpha}_{\alpha}(r)- g^{-1}(r)H_{2}(r) & 0 \\
 0 &  g^{-1}(r)H_{2}(r)
\end{array}
\right]+ g^{-1}(r) \left[ \begin{array}{cc}
-\beta & -\delta\\
\delta & \beta
\end{array}
\right] \label{generalT} \ee or,  substituting (\ref{dconstant})
and (\ref{H21}), a more explicit expression is \bea T^{\mu}_{\nu}=
& \left[
\begin{array}{cc}
 \frac{\Lambda}{12\pi}+\frac{J^{2}}{16r^{4}}
-\frac{1}{96\pi g(r)}\left[2\Lambda r -\frac{J^{2}}{2
r^{3}}\right]^{2}+ g^{-1}(r)D
& 0 \\
 0 & \frac{1}{96\pi g(r)}\left[2\Lambda r -\frac{J^{2}}{2
r^{3}}\right]^{2} - g^{-1}(r)D
\end{array} \right] & \nn\\
 & +\hspace{0.5cm} g^{-1}(r)
\left[ \begin{array}{cc}
-\beta & -\alpha g^{\frac{3}{2}}(r) e^{-\frac{1}{4} g^{2}(r)} \\
\alpha g^{\frac{3}{2}}(r) e^{-\frac{1}{4} g^{2}(r)}  & \beta
\end{array}
\right]
 \label{em3} \eea where the Achucarro-Ortiz black hole background
(\ref{metric1})-(\ref{metric2}) and relations (\ref{aotrace}),
(\ref{dconstant}), (\ref{H21}) and (\ref{constant}) have been
used. In this expression the only unknowns are the parameters
$\alpha$ and $\beta$; we hope to determine them imposing the third
and fourth Wald's axioms treating the Achucarro-Ortiz black hole
as a Casimir system \cite{birrell}. Two one-dimensional ``walls"
at a proper distance (between them) $L$ are placed at
points\hspace{1ex} $r_1$ and $r_2$. The massless scalar field
whose energy-momentum tensor we try to evaluate satisfies the
Dirichlet boundary conditions on the ``walls'', i.e.
$\phi(r_{1})=\phi(r_{2})=0$.
\par\noindent
We are now going to find the explicit form of the regularized
stress tensor in the different vacua.
\newline
\newline
\noindent {\bf (i) Boulware Vacuum}
\newline
In this vacuum there are no particles detected at infinity
($\mathcal{J}^+$) and the regularized stress tensor (\ref{em3})
should coincide at infinity with the sum of the standard Casimir
stress tensor \cite{birrell,fulling} in the Minkowski spacetime
\be T^{\mu}_{\nu}=\frac{\pi}{24L^{2}} \left[\begin{array}{cc}
-1 & 0 \\
0 & 1
\end{array}
\right] \label{emb} \ee and of the stress tensor of the
gravitational field of the AdS$_{2}$ space \bea T^{\mu
(\mbox{\tiny{AdS}})}_{\nu}&=& \left[\begin{array}{cc}
\displaystyle{-\frac{r^2}{2\left(-M+\Lambda r^{2}
\right)}}&\displaystyle{0}
\\ \displaystyle{0}&\displaystyle{\frac{r^2}{2\left(-M+\Lambda r^2
\right)^2}}\end{array}\right]\nn\\
&=&\frac{r^{2}}{2\left(-M+\Lambda r^{2}
\right)}\left[\begin{array}{cc} -1 & 0 \\
0 & 1\end{array}\right]\label{adsstress1}\eea since the
Achucarro-Ortiz black hole is asymptotically an AdS$_{2}$ space.
\newline The constants of integration $\alpha$ and $\beta$ are
evaluated demanding the regularized stress tensor given in
(\ref{em3}) to coincide at infinity, i.e. $r\rightarrow +\infty$,
with the sum of the above-mentioned stress tensors (\ref{emb}) and
(\ref{adsstress1}).
\par
\noindent Therefore we get  \be \alpha = 0 \hspace{1cm}\beta =
\left(\frac{\Lambda}{12\pi}+\frac{\pi}{24L^{2}}\right)
g_{\mbox{\tiny{AdS}}}(r)
+\left(\frac{1}{2}-\frac{\Lambda^{2}}{24\pi}\right)r^{2}+D
 \ee and the regularized stress tensor has
been explicitly calculated. It can also be written as a direct sum
\be
T^{(\eta)\mu}_{\nu}=T^{\mu}_{\nu(gravitational)}+T^{\mu}_{\nu(boundary)}+T^{\mu}_{\nu(ANFG)}
\label{sum1}\ee where $\eta$ denotes that the regularized stress
tensor has been calculated under the assumption that there are no
particles (vacuum state) at infinity (Boulware vacuum). The first
term denotes the contribution to the vacuum polarization due to
the non-trivial topology in which the contribution of the trace
anomaly is included, the second term denotes the contribution due
to the presence of the two ``Dirichlet walls'' and the third term
denotes the contribution due to the asymptotically non-flat
geometry (ANFG) of the Achucarro-Ortiz black hole.
\par
The detected energy density, pressure and energy at infinity
($r\rightarrow +\infty$ are given by
 \bea \rho=T^{(\eta)t}_{t}&=&-\frac{\pi}{24L^{2}}-\frac{1}{2\Lambda}\label{d1}\\
p=-T^{(\eta)x}_{x}&=&-\frac{\pi}{24L^{2}}-\frac{1}{2\Lambda}-\frac{\Lambda}{12\pi}\label{p1}\\
E(L)=\int^{r_2 =r_1 +L}_{r_1}\rho dr &=&
-\frac{\pi}{24L}-\frac{1}{2\Lambda}L\label{energy1}\hspace{1ex}.
\eea The corresponding Casimir force between the ``walls'' is not
always attractive as expected  \be F(L)=-\frac{\partial
E(L)}{\partial
L}=-\frac{\pi}{24L^{2}}+\frac{1}{2\Lambda}\label{f1}\hspace{1ex}.
\ee It is clear that the Casimir force is
\begin{description}
\item \hspace{2cm}{\bf (a) attractive}
\be L < \sqrt{\frac{\pi}{12}}\,\Lambda^{1/2} \ee
\item \hspace{2cm}{\bf (b) zero} \be L = \sqrt{\frac{\pi}{12}}\,\Lambda^{1/2} \ee
\item \hspace{2cm}{\bf (c) repulsive}
\be L > \sqrt{\frac{\pi}{12}}\,\Lambda^{1/2}\hspace{1ex}.\ee
\end{description}
\mbox{}
\\
\\
\par\noindent {\bf (ii) Hartle-Hawking Vacuum}
\newline
In this vacuum the Achucarro-Ortiz black hole
(\ref{metric1})-(\ref{metric2}) is in thermal equilibrium with an
infinite reservoir of black body radiation at temperature $T$
which is equal to its Hawking temperature. The regularized stress
tensor (\ref{em3}) should coincide with the following stress
tensor  \be T^{\mu}_{\nu}=\frac{\pi}{24L^{2}}
\left[\begin{array}{cc}
-1 & 0 \\
0 & 1
\end{array}
\right]+\frac{r^{2}}{2\left(-M+\Lambda r^{2}
\right)}\left[\begin{array}{cc} -1 & 0 \\
0 & 1\end{array}\right]+\frac{\pi T^{2}}{6} \left[
\begin{array}{cc}
-1 & 0\\
0 & 1
\end{array}
\right ]\hspace{1ex} \ee  where the last term is the stress tensor
for a two-dimensional black hole in thermal equilibrium at
temperature $T$ \cite{christensen}.
\newline The constants of integration $\alpha$ and $\beta$ are
evaluated demanding the regularized stress tensor given in
(\ref{em3}) to coincide at infinity, i.e. $r\rightarrow +\infty$,
with the sum of the above-mentioned stress tensors (\ref{emb}) and
(\ref{adsstress1}).
\par
\noindent Therefore we get  \bea \alpha &=& 0 \\
\beta
&=&\left(\frac{\Lambda}{12\pi}+\frac{\pi}{24L^{2}}+\frac{\pi}{6}
\left(T_{H}^{\mbox{\tiny{AdS}}}\right)^{2} \right)
g_{\mbox{\tiny{AdS}}}(r)
+\left(\frac{1}{2}-\frac{\Lambda^{2}}{24\pi}\right)r^{2}+D\\
&=& \left(\frac{\Lambda}{12\pi}+\frac{\pi}{24L^{2}}+\frac{\Lambda
}{24\pi}M\right) g_{\mbox{\tiny{AdS}}}(r)
+\left(\frac{1}{2}-\frac{\Lambda^{2}}{24\pi}\right)r^{2}+D  \eea
and the regularized energy-momentum tensor has been explicitly
calculated. It can also be written as a direct sum \be
T^{(\upsilon)\mu}_{\nu}=T^{\mu}_{\nu(gravitational)}+T^{\mu}_{\nu(boundary)}+
T^{\mu}_{\nu(ANFG)}+T^{\mu}_{\nu(bath)} \label{sum2}\ee where
$\upsilon$ denotes that the regularized stress tensor has been
calculated under the assumption that massless particles (black
body radiation) are detected at infinity (towards
$\mathcal{J}^{+}$) (Hartle-Hawking vacuum) and the fourth term in
equation (\ref{sum2}) denotes the contribution to the vacuum
polarization due to the thermal bath at temperature $T_{H}$.
\par
In this vacuum the asymptotically ($r\rightarrow +\infty$)
detected energy density, pressure and energy at infinity are given
by
 \bea \rho=T^{(\eta)t}_{t}&=&-\frac{\pi}{24L^{2}}-\frac{1}{2\Lambda}-\frac{\Lambda}{24\pi}M\label{d2}\\
p=-T^{(\eta)x}_{x}&=&-\frac{\pi}{24L^{2}}-\frac{1}{2\Lambda}-\frac{\Lambda}{24\pi}M-\frac{\Lambda}{12\pi}\label{p2}\\
E(L)=\int^{r_2 =r_1 +L}_{r_1}\rho dr &=&
-\frac{\pi}{24L}-\frac{1}{2\Lambda}L -\frac{\Lambda }{24\pi}M
L\label{energy2}\hspace{1ex}. \eea The corresponding Casimir force
between the ``walls'' is not always attractive as expected : \be
F(L)=-\frac{\partial E(L)}{\partial
L}=-\frac{\pi}{24L^{2}}+\frac{1}{2\Lambda}+\frac{\Lambda}{24\pi}M\label{f2}\hspace{1ex}.
\ee It is clear that the Casimir force is :
\begin{description}
\item \hspace{2cm}{\bf (a) attractive}
\be L < \pi \sqrt{\frac{\Lambda}{12\pi +\Lambda^{2} M }}\ee
\item \hspace{2cm}{\bf (b) zero} \be L = \pi \sqrt{\frac{\Lambda}{12\pi +\Lambda^{2} M }}\ee
\item \hspace{2cm}{\bf (c) repulsive}
\be L > \pi \sqrt{\frac{\Lambda}{12\pi +\Lambda^{2} M
}}\hspace{1ex}.\ee
\end{description}
Thus, if the last condition is satisfied the outer ``Dirichlet
wall'' moves towards infinity. It can be studied as a ``moving
mirror'' creating particles whose energy rate detected at infinity
is given by the third term in equation (\ref{energy2}): \be
\frac{dE}{dt}=\frac{\Lambda}{24 \pi}M L =\frac{\pi
L}{6}\left(T_{H}^{\mbox{\tiny{AdS}}}\right)^{2}\hspace{1ex}. \ee
This is the rate at which energy is radiated for the case of the
massless two-dimensional field \cite{emparan1,emparan2}.
\newline
\newline
\par
\noindent {\bf (iii) Unruh Vacuum}
\newline
In this vacuum an outward flux of radiation is detected at
infinity. Thus, since the Achucarro-Ortiz black hole
(\ref{metric1})-(\ref{metric2}) radiates and its spectrum
distribution is thermal at the Hawking temperature $T_{H}$
\cite{hawking2,hawking3}, the Unruh vacuum state is identified
with the vacuum obtained after the Achucarro-Ortiz black hole has
settled down to an ``equilibrium'' of temperature $T_{H}$. The
regularized stress tensor (\ref{em3}) should now coincide at
infinity with the following stress tensor \be
T^{\mu}_{\nu}=\frac{\pi}{24L^{2}} \left[\begin{array}{cc}
-1 & 0 \\
0 & 1
\end{array}
\right]+\frac{r^{2}}{2\left(-M+\Lambda r^{2}
\right)}\left[\begin{array}{cc} -1 & 0 \\
0 & 1\end{array}\right]+\frac{\pi
\left(T_{H}^{\mbox{\tiny{AdS}}}\right)^{2} }{12} \left[
\begin{array}{cc}
-1 & -1\\
1 & 1
\end{array}
\right ]\hspace{1ex} \ee  where the last term is the stress tensor
for a radiating two-dimensional black hole which has settled down
to an ``equilibrium'' of temperature $T_{H}$ \cite{christensen}.
\newline The constants of integration $\alpha$ and $\beta$ are
evaluated demanding the regularized stress tensor given in
(\ref{em3}) to coincide at infinity, i.e. $r\rightarrow +\infty$,
with the sum of the above-mentioned stress tensors (\ref{emb}) and
(\ref{adsstress1}).
\par
\noindent Therefore we get  \bea \alpha &=& \frac{\pi
\left(T_{H}^{\mbox{\tiny{AdS}}}\right)^{2}
}{12}g^{-\frac{1}{2}}_{\mbox{\tiny{AdS}}}(r)e^{\frac{1}{4}g^{2}_{\mbox{\tiny{AdS}}}(r)}
=\frac{\Lambda
}{48\pi}M g^{-\frac{1}{2}}_{\mbox{\tiny{AdS}}}(r)e^{\frac{1}{4}g^{2}_{\mbox{\tiny{AdS}}}(r)}\\
\beta
&=&\left(\frac{\Lambda}{12\pi}+\frac{\pi}{24L^{2}}+\frac{\pi}{12}
\left(T_{H}^{\mbox{\tiny{AdS}}}\right)^{2} \right)
g_{\mbox{\tiny{AdS}}}(r)
+\left(\frac{1}{2}-\frac{\Lambda^{2}}{24\pi}\right)r^{2}+D\nn\\
&=& \left(\frac{\Lambda}{12\pi}+\frac{\pi}{24L^{2}}+\frac{\Lambda
}{48\pi}M\right) g_{\mbox{\tiny{AdS}}}(r)
+\left(\frac{1}{2}-\frac{\Lambda^{2}}{24\pi}\right)r^{2}+D  \eea
and the regularized stress tensor has been explicitly calculated.
It can also be written as a direct sum \be
T^{(\xi)\mu}_{\nu}=T^{\mu}_{\nu(gravitational)}+T^{\mu}_{\nu(boundary)}+
T^{\mu}_{\nu(ANFG)}+T^{\mu}_{\nu(radiation)} \label{sum3}\ee where
$\xi$ denotes that the regularized stress tensor has been
calculated under the assumption that massless particles are
detected at infinity due to the Hawking radiation of the
Achucarro-Ortiz black hole (Unruuh vacuum) and the fourth term in
equation (\ref{sum3}) denotes the contribution to the vacuum
polarization due to Hawking radiation at temperature $T_{H}$.
\par
In this vacuum the asymptotically ($r\rightarrow +\infty$)
detected energy density, pressure and energy at infinity are given
by
 \bea \rho=T^{(\eta)t}_{t}&=&-\frac{\pi}{24L^{2}}-\frac{1}{2\Lambda}-\frac{\Lambda}{48\pi}M\label{d3}\\
p=-T^{(\eta)x}_{x}&=&-\frac{\pi}{24L^{2}}-\frac{1}{2\Lambda}-\frac{\Lambda}{48\pi}M-\frac{\Lambda}{12\pi}\label{p3}\\
E(L)=\int^{r_2 =r_1 +L}_{r_1}\rho dr &=&
-\frac{\pi}{24L}-\frac{1}{2\Lambda}L -\frac{\Lambda }{48\pi}M
L\label{energy3}\hspace{1ex}. \eea The corresponding Casimir force
between the ``walls'' is not always attractive as expected : \be
F(L)=-\frac{\partial E(L)}{\partial
L}=-\frac{\pi}{24L^{2}}+\frac{1}{2\Lambda}+\frac{\Lambda}{48\pi}M\label{f3}\hspace{1ex}.
\ee It is clear that the Casimir force is :
\begin{description}
\item \hspace{2cm}{\bf (a) attractive}
\be L < \pi \sqrt{\frac{2\Lambda}{24\pi +\Lambda^{2} M }}\ee
\item \hspace{2cm}{\bf (b) zero} \be L = \pi \sqrt{\frac{2\Lambda}{24\pi +\Lambda^{2} M }}\ee
\item \hspace{2cm}{\bf (c) repulsive}
\be L > \pi \sqrt{\frac{2\Lambda}{24\pi +\Lambda^{2} M
}}\hspace{1ex}.\ee
\end{description}
As in the case of the Hartle-Hawking vacuum, if the last condition
is satisfied the outer ``wall'' moves towards infinity. It can be
studied as a ``moving mirror'' creating particles whose energy
rate detected at infinity is given by the second term in equation
(\ref{energy3}): \be \frac{dE}{dt}=\frac{\Lambda}{48 \pi}M L
=\frac{\pi
L}{12}\left(T_{H}^{\mbox{\tiny{AdS}}}\right)^{2}\hspace{1ex}. \ee
This is the rate at which energy is radiated for the case of the
massless two-dimensional field \cite{emparan1,emparan2}.
\section{Equilibrium and Cosmological Constant}
It is obvious that in the case that the net force which the
``Dirichlet walls'' experience turns out to be repulsive the
system will be uninteresting since it will be decompactified as
$L\rightarrow \infty$. On the other hand, if the net forced
exerted on the ``Dirichlet walls'' turns out to be attractive then
the system inevitably will evolve in such a way that at some
finite time the distance $L$ will be of order of the Planck length
where our semiclassical analysis adopted here will no longer be
valid. Therefore the case of a zero net force on the ``Dirichlet
walls'' sounds the most interesting for our scenario.
\par\noindent
The net force exerted on the ``Dirichlet walls'' can be evaluated
using the Casimir force in any of the three vacua. It should be
noted that for the cases of the Hartle-Hawking and Unruuh vacua
the last term in equations (\ref{f2}) and (\ref{f3}),
respectively, should be removed. The reason is that in both vacua
the forces acting on both sides of each ``Dirichlet wall'' due to
thermal bath or radiation , respectively, are the same and thus
their total contribution to the net force is zero. Therefore the
net force which the ``Dirichlet walls'' experience is given as \be
F_{net}=-\frac{\pi}{24L^2}+\frac{1}{2\Lambda}\ee and setting it
equal to zero the distance $L$ between the ``Dirichlet walls''
receives the value \be
L=\sqrt{\frac{\pi}{12}}\Lambda^{1/2}\hspace{1ex}.\ee It is clear
that the distance $L$ between the ``Dirichlet walls'' is
controlled by the value of the cosmological constant.
\section{Conclusions}
In this paper we have explicitly calculated in the Achucarro-Ortiz
black hole background the regularized stress tensor of a massless
scalar field satisfying Dirichlet boundary conditions on
one-dimensional ``walls'' (``Dirichlet walls''). The regularized
stress tensor is separately treated in the Boulware,
Hartle-Hawking and Unruh vacua. In all these vacua, expressions
for the asymptotically detected energy, energy density and
pressure acting on the ``Dirichlet walls'' are obtained. The
values of the above mentioned quantities are all negative
exhibiting a violation of all energy conditions \cite{hawking4}.
This ``problem'' is somehow expected to take place in our scenario
since violations of some or all of the energy conditions appear as
soon as scalar fields couple to gravity \cite{matt1,matt2}. In the
Hartle-Hawking and Unruuh vacua, the corresponding Casimir force
is evaluated and proved, as expected, to be not always attractive:
it can be attractive, repulsive or zero according to the distance
$L$ between the ``Dirichlet walls''. In contradistinction to what
is known till now \cite{setare1,setare2,setare3,elias}, in the
Boulware vacuum, the Casimir force is also not always attractive.
Additionally, we have evaluated the net force exerted on the
``Dirichlet walls''. It has been easily demonstrated by imposing
the condition of equilibrium on the ``Dirichlet walls'', i.e. zero
net force, that the distance between the one-dimensional ``walls''
is tuned by the cosmological constant\footnote{The stability of
the configurations can be checked by using equation
(\ref{energy1}) in the Boulware vacuum. It is easily derived that
the configurations are unstable against small displacements. Same
result can be derived by using equations (\ref{energy2}) and
(\ref{energy3}) in the Hartle-Hawking and Unruuh vacua,
respectively, but the last term in these equations has to be
dropped for the reason given in Section 6.}.
\par
It would be very interesting for our scenario to be utilized in
higher dimensions and specifically in braneworlds. Of course, it
is well-known that the trace anomaly -which plays a key role in
the technique presented here-  is zero for odd-dimensional
spacetimes. Therefore, only even-dimensional spacetimes should be
considered. It should also be pointed out that our scenario is not
directly applicable to higher even-dimensional spacetimes, since
more conditions are required in order to completely fix the form
of the renormalized stress tensor corresponding to the quantized
scalar field. Indeed, there are a number of recent works which
they deal with the Casimir effect in different models of
braneworlds \cite{garriga1,kanti,sergei1,maroto,sergei2,garriga2}.
These scenarios are more complicated than the one analyzed here,
just to mention that the branes are located in the bulk space
\cite{rs1,rs2}, not at points of the spacetime in which we live,
or the existence of the radion field which has to be stabilized
\cite{wise}.
\section*{Acknowledgements}
 The author is grateful to Professor R. Wald and Associate
 Professor N. Tetradis for useful correspondences.
 The author is  also indebted to Professor J. Garriga and  Assistant Professor T. Christodoulakis
for valuable discussions and enlightening comments on a draft of
this paper. Many thanks are acknowledged to the referee for
helpful and fruitful comments. This work has been supported by the
European Research and Training Network ``EUROGRID-Discrete Random
Geometries: from Solid State Physics to Quantum Gravity"
(HPRN-CT-1999-00161).



\begin{thebibliography}{99}


\bibitem{birrell}  N.D. Birrell and P.C.W. Davies, {\it Quantum fields in curved space}
(Cambridge University Press, 1982).

\bibitem{fulling}  S.A. Fulling, {\it Aspects of Quantum Field Theory in
Curved Space-Time} (Cambridge University Press, 1989).

\bibitem{rob} R.M. Wald, {\it Quantum Field Theory in Curved Spacetime and Black Hole
Thermodynamics},(The University of Chicago Press, 1994).

\bibitem{bryce} B.S. DeWitt, Phys. Rep. {\bf C 19} 295 (1975).

\bibitem{kummer} W. Kummer and D.V. Vassilevich, Ann. Phys. (Leipzig) {\bf 8}: 801 (1999) and references therein.

\bibitem{capper1} D.M. Capper and M.J. Duff, Nuovo Cimento A {\bf23} 173 (1974).

\bibitem{capper2} D.M. Capper and M.J. Duff, Phys. Lett. A {\bf 53} 361 (1975).

\bibitem{deser} S. Deser, M.J. Duff and C.J. Isham, Nucl. Phys. B {\bf 111} 45 (1976).

\bibitem{plunien} G.Plunien, B. Muller and W. Greiner, Phys. Rep. {\bf134} 87 (1986).

\bibitem{candelas} D. Deutsch, and  P. Candelas, Phys. Rev. {\bf D 20} 3063 (1979).

\bibitem{greiner} P. Greiner, Arch. Rat. Mech. Anal. {\bf41} 163 (1971).

\bibitem{gilkey} P.B. Gilkey, Comp. Math. {\bf38} 201 (1979).

\bibitem{hawking1} S.W. Hawking, Commun. Math. Phys. {\bf55} 133 (1977).

\bibitem{davies1} P.C.W. Davies, S.A. Fulling, S.M. Christensen and T.S. Bunch, Ann. Phys. {\bf109} 108 (1977).

\bibitem{davies2} T.S. Bunch and  P.C.W. Davies,  Proc. Roy. Soc. London,  Sect A {\bf 360} 117 (1978).

\bibitem{christ} S.M. Christensen, Phys. Rev. {\bf D 14} 2490 (1976).

\bibitem{vilenkin} A. Vilenkin, Nuovo Cimento A {\bf 44} 441 (1978).

\bibitem{christensen} S.M. Christensen and  S.A. Fulling, Phys. Rev. {\bf D 15} 2088 (1977)
 and references therein.

\bibitem{wald1} R.M. Wald, Commun. Math. Phys. {\bf54} 1 (1977).

\bibitem{wald2} R.M. Wald, Phys. Rev.  {\bf D 17} 1477 (1978).

\bibitem{casimir}  H.B.G. Casimir, Konink. Nederl. Akad. Weten., Proc. Sec. Sci. {\bf51} 793 (1948).

\bibitem{bressi} G. Bressi, G. Carugno, R. Onofrio and G. Ruoso,
Phys. Rev. Lett. {\bf88} 041804 (2002).

\bibitem{bordag} M. Bordag, U. Mohideen and V.M. Mostepanenko,
Phys. Rep. {\bf353} 1 (2001).

\bibitem{mostepanenko} V.M. Mostepanenko and  N.N. Trunov, {\it The Casimir Effect
 and its Applications}, (Clarendon Press, Oxford, 1997).

\bibitem{emilio1}  E. Elizalde, Phys. Lett. B {\bf 213} 477 (1988).

\bibitem{emilio2} M. Bordag, E. Elizalde, K. Kirsten and   S. Leseduarte, Phys. Rev. {\bf D 56} 4896 (1997) .

\bibitem{emilio3}  E. Elizalde, Z. Phys. {\bf C 44} 471 (1989).

\bibitem{ana} Ana Ach\'{u}carro and M. Ortiz, Phys. Rev. D {\bf48} 3600
(1993).

\bibitem{martinez} D. Louis-Martinez and G. Kunstatter, Phys. Rev. D
{\bf52} 3494 (1995).

\bibitem{setare1} M.R. Setare and  A.H. Rezaeian, Mod. Phys. Lett. A {\bf 15} 2159 (2000).

\bibitem{setare2} M.R. Setare, Class. Quant. Grav. {\bf18} 2097 (2001).

\bibitem{setare3} M.R. Setare, {\it Trace Anomaly and Backreaction of the Dynamical Casimir
Effect}, hep-th/0205081.

\bibitem{elias} T. Christodoulakis, G.A. Diamandis, B.C. Georgalas and  E.C. Vagenas, Phys. Rev. D
{\bf64} 124022 (2001).

\bibitem{boulware} D.G. Boulware, Phys. Rev. {\bf D 13} 2169 (1976).

\bibitem{hartle} J.B. Hartle and  S.W. Hawking, Phys. Rev. {\bf D 13} 2188 (1976).

\bibitem{gibbons}  G.W. Gibbons and  M.J. Perry, Phys. Rev. Lett. {\bf36} 985 (1976).

\bibitem{israel} W. Israel, Phys. Lett. A {\bf 57} 107 (1976).

\bibitem{unruh} W.G. Unruh, Phys. Rev. {\bf D 14} 870 (1976).

\bibitem{wald3} S. Hollands and  R.M. Wald, Comm. Math. Phys. {\bf223}
289 (2001).

\bibitem{capri} A.Z. Capri, M. Kobayashi and  D.J. Lamb, Class. Quant. Grav. {\bf13} 179 (1996).

\bibitem{jorge} M. Ba\~{n}ados, C. Teitelboim and  J. Zanelli, Phys.
Rev. Lett. {\bf69} 1849 (1992).


\bibitem{kumar1} A. Kumar and  K. Ray, Phys. Lett. B {\bf351} 431
(1995).

\bibitem{kim} W.T. Kim and J.J. Oh, Phys. Lett. B {\bf461} 189 (1999).

\bibitem{emparan1} R. Emparan, G.T. Horowitz and  R.C. Myers, Phys.
Rev. Lett. {\bf85} 499 (2000).

\bibitem{emparan2} R. Emparan, hep-th/0009136. in  Proceedings of the
4th Annual European TMR Conference on Integrability. Non
Perturbative Effects and Symmetry in Quantum Field Theory, Paris,
France, 7-13 Sep. 2000.

\bibitem{hawking2} S.W. Hawking, Nature {\bf248} 30 (1974).

\bibitem{hawking3} S.W. Hawking, Commun. Math. Phys. {\bf43} 199 (1975).

\bibitem{hawking4} S.W. Hawking and  G.F.R. Ellis, {\it The large scale structure of the
spacetime}, (Cambridge University Press, 1973).

\bibitem{matt1} C. Barcel\mbox{\'{o}} and  M. Visser, Class. Quant.
Grav. {\bf17} 3843 (2000).

\bibitem{matt2} C. Barcel\mbox{\'{o}} and  M. Visser, Int. J. Mod. Phys. D {\bf171} 1553 (2002).

\bibitem{garriga1} J. Garriga, O. Pujolas and T. Tanaka, Nucl. Phys.
B {\bf605} 192 (2001).

\bibitem{kanti} R. Hofmann, P. Kanti and  M. Pospelov, Phys. Rev. D
{\bf63} 124020 (2001).

\bibitem{sergei1} S. Nojiri, S.D. Odintsov and  A. Sugamoto, Mod.
Phys. Lett. A {\bf17} 1269 (2002).

\bibitem{maroto} A.L. Maroto, Nucl. Phys. B {\bf653} 109 (2003).

\bibitem{sergei2} E.Elizalde, S. Nojiri, S.D. Odintsov and  S.
Ogushi, Phys. Rev. D {\bf67} 063515 (2003).

\bibitem{garriga2} J. Garriga and  A. Pomarol, Phys. Lett. B {\bf560} 91 (2003).

\bibitem{rs1} L. Randall and  R. Sundrum, Phys. Rev. Lett. {\bf83}
3370 (1999).

\bibitem{rs2} L. Randall and  R. Sundrum, Phys. Rev. Lett. {\bf83}
4690 (1999).

\bibitem{wise}W.D. Goldberger and M.B. Wise, Phys.Rev. Lett. {\bf83}
4922 (1999).


\end{thebibliography}
\end{document}